# Towards the Interactive Effects of Demand Response Participation on Electricity Spot Market Price

Saeed Mohajeryami, Milad Doostan, Seyedmahdi Moghadasi
Energy Production and Infrastructure Center (EPIC)
Electrical and Computer Engineering Department
University of North Carolina at Charlotte
Charlotte, NC, USA
Email:{smohajer,mdoostan,smoghada}@uncc.edu

Peter Schwarz
Belk College of Business
University of North Carolina at Charlotte
Charlotte, NC, USA
Email: pschwarz@uncc.edu

*Abstract*—The electricity market is threatened by supply scarcity, which may lead to very sharp price spikes in the spot market. On the other hand, demand-side's activities could effectively mitigate the supply scarcity and absorb most of these shocks and therefore smooth out the price volatility. In this paper, the positive effects of employing demand response programs on the spot market price are investigated. A demand-price elasticity based model is used to simulate the customer reaction function in the presence of a real time pricing. The demand achieve by DR program is used to adjust the spot market price by using a price regression model. SAS software is used to run the multiple linear regression model and MATLAB is used to simulate the demand response model. The approach is applied on one week data in summer 2014 of Connecticut in New England ISO. It could be concluded from the results of this study that applying DR program smooths out most of the price spikes in the electricity spot market and considerably reduces the customers' electricity cost.

## I. Introduction

The market for electricity are progressively developing over numerous years of competition and reorganization. Nevertheless, there are still several areas in the industry that are kept shielded from the advancement in the market, which one of them is demand-side. As a matter of fact, this area is underdeveloped due to the detachment from market price fluctuations as the regulatory bodies attempted to give immunity to retail customers vulnerable to such fluctuations. For instance, during 2000 and 2001, California experienced a major power crisis under the restructured wholesale market. Although numerous factors could be listed as reasons to create this crisis, most people agree that the lack of demand response exacerbated the situation[1].

However, studies over the recent years show that demand response (DR) programs could create an environment in which customers could engage in the process of optimized decision making. Consequently, it can change the customers' consumption pattern in response to the price signals provided by the wholesale market.

These programs can create numerous practical possibilities for the power system operators and utilities to make an improvement in both economic and technical indices of their system. As a matter of fact, power system operators can compensate lack of their supply during the peak time with DR resources. It is estimated that the capacity to meet demand during the top 100 peak hours accounts for 10-20% of electricity cost annually [2]. On the other hand, utilities can also benefit from DR by taking advantages of lower prices offered by such resources compared to electricity spot market.

DR programs are generally could be separated into two main categories: incentive-based programs (IBP) and time-based rate (TBR) programs. As shown in Fig. 1, each category is composed of several programs. The authors in [3-4] elaborated these programs in detail.

For many years, utilities have offered IBPs to to large industrial and commercial customers. As an example, ERCOT offers emergency interruptible load program to its large customers. Moreover, Southern California Edison (SCE) has offered a variety of DR programs such as automated demand response (Auto-DR), permanent load shifting (PLS) and scheduled load reduction programs (SLRP) to its large customers [5]. On the other hand, TBR programs are typically neglected by utilities due to the lack of proper infrastructure, technical complication and highly capital intensive infrastructure. However, over the last decade, the US government and its energy sector, due to environmental challenges of the traditional electricity generation, attempted to adopte a supportive approach in order to provide the necessary infrastructure for DR and energy efficiency programs. These programs are mostly TBR programs. Another impact factor that helps to launch TBR programs more conveniently is the penetration of advanced smart meters. Advanced metering penetration, based on 2010 FERC survey, has reached to a considerable level of 8.7 percent in the US [6]. Advanced metering is regarded by many as a cornerstone of the TBR programs. Furthermore, many utilities have recently launched pilot programs to evaluate the feasibility and technical challenges of TBR programs in the new environment [7-11].

Although based on the previous discussion, the financial support is more available in order to build the necessary infrastructure for the TBR programs, these program face many obstacles in the implementation stage. One of the most major issues that utilities face with regard to the design of the programs is finding the proper model to explain the customer's reaction to the incentives provided by each program. As a matter of fact, the utilities cannot employ the proper profit maximizing strategies in the absence of a reliable model. Therefore, many of the programs might not even initiated or if they do, they might be doomed to failure. Moreover, the wrong models might leads to proposing the wrong incentive payments. The improper incentive payment can discourage the participation in the program. Therefore, to overcome this problem, different models are proposed to explain the customer's reaction function. Authors in [12-14] employ demand-price elasticity concept to model the effects of implementing demand response programs on customer's reaction. In fact, demand-price elasticity is a concept borrowed from the consumer theory in microeconomics which reflects the relative change in the demand with respect to the relative change in the price [15-16]. The approach utilized in [17] models the customer's reaction function with linear optimization technique assuming the customers have access to the real-time electricity prices. In this model, the objective function is maximizing the utility and minimizing the cost of electricity consumption.

A statistical method introduced in [18] uses the demand-price elasticity to explain the customer reaction function in the direct load control (DLC) program. Moreover, the authors in [19] applies self-organizing maps and statistical Ward's linkage to classify electricity market prices into different clusters. It also uses a non-parametric curve estimation approach to explore the underlying structure in different clusters which leads to extraction of the proper customer's reaction to the different prices. Furthermore, the authors in [20] developed a method based on consumers theory in microeconomics to incorporate the customers' willingness to shift consumption cross-periods based on the pertinent rates.

In addition to the aforementioned models, several forecasting based approaches are proposed to explain the customer's reaction function. These approaches which use ex-post and ex-ante data are being used to forecast the short-term and long-term customer's reaction function. Authors in [21] report the current practice at Pacific Gas and Electric (PG&E) company. ex-ante and ex-post reports are used to develop individual customer regression models. In order to develop a robust model, all the interdependencies of weather, calendar days, etc are added to the regression model. The utilized models need substantial amount of historical data and proper control groups.

In this paper, the effect of demand response on the spot market price is examined. The objective is to investigate the impact of DR on the price volatility by proposing an algorithm which feeds the outcome of DR model into the spot market and explore the positive impact of demand response. It will be shown how the full participation of the demand side under real time pricing can decrease the wholesale price and its fluctuations in the market. In order to carry out such task, the customer's reaction function model and day-ahead load forecasting are required. In this paper, customer's reaction model is taken from [22]; also, for the forecasting part, multiple linear regression model (MLR) is used.

Two main classes of load and price forecasting are prevalent in the literature. One assumes that merely the availability of the historical data of the desired variable is sufficient for forecast purposes, while the other relies on additional different parameters like weather, pressure, humidity, seasonality, etc to do the forecasting. ARIMA models which belong to the former class are popular in short term load and price forecasting [23-24]. The latter class which uses MLR models are suitable for short, medium and long term forecasting. Both classes could be enhanced by different intelligent techniques such as Artificial Neural Network (ANN), Fuzzy logic and Wavelet [25-28]. [29] analyzes the application of aforementioned load forecasting classes in the presence of DR programs in detail.

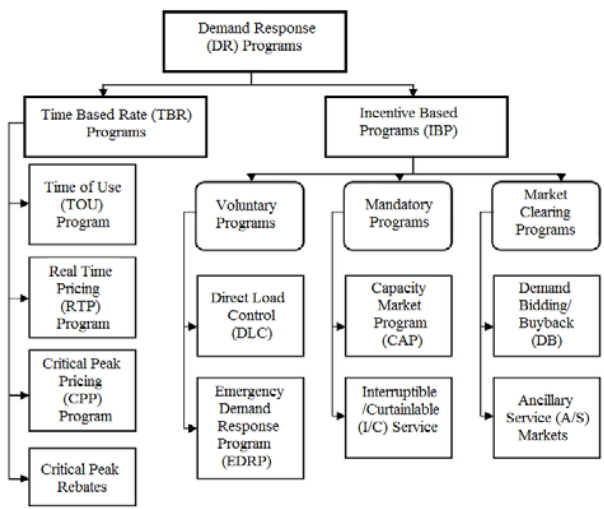

Fig. 1. Categories of the demand response programs

Even though, numerous models are proposed for price forecasting, there is not any single model that works for all the situations; consequently, the utilities use multiple models in parallel to create scenario based forecasts.

The organization of this paper is as follows. First, a full description of the demand response model used in this paper is presented in section II. Multiple linear regression model is elaborated in Section III. Section IV explains the implementation of the proposed algorithm. Then the results for the case study as well as discussion of the results are provided in section V. Section VI closes the paper with drawing conclusion from the provided discussion and results.

## II. DEMAND RESPONSE MODEL

In order to describe the employed demand response model in this paper, it is necessary to understand the concept of demand-price elasticity. The demand for almost all goods and services rises as the price decreases. Based on diminishing marginal return law, this change in the demand is not linear [15]. In order to quantify the aforesaid change in the demand, the concept of demand-price elasticity has to be utilized. Indeed, the nonlinear demand curve could be linearized around a given point. Then the change in the demand relative to the change in price could be measured which is known as demand-price elasticity. (1) represents the demand-price elasticity function mathematically.

$$E = \frac{P_0}{d_0} \times \frac{\partial d}{\partial p} \qquad (1)$$

Where $E$ is the price elasticity, $p$ and $d$ are price and demand, $P_0$ and $d_0$ are initial price and demand respectively.

Price elasticity has two components: self-elasticity and cross-elasticity. In other words, between two competing commodities, the percent change of demand with respect to the percent change in its own price is self-elasticity, whereas the percent change of demand with respect to the percent change in the price of the other commodity is cross-elasticity.

As it was mentioned, the demand response model in this paper is defined based on the elasticity. To achieve this target demand response model for 24 hours, first it is required that the model for one hour to be extracted and then expanded to 24 hours. In what follows, this procedure is described.

*A. Demand-price elasticity model for one hour*

Suppose the customer's benefit for the *i*-th hour is as follows:

$$B(d(i)) = U(d(i)) - d(i).p(i) \quad (2)$$

Where $U(d(i))$ is customer's utility in *i*-th hour. This function could be formulated with the Taylor serious expansion accourding to [29].

$$U(d(i)) = U(d_0(i)) + \frac{\partial U(d_0(i))}{\partial d(i)} \times \Delta d(i) + \quad (3)$$

$$\frac{1}{2} \times \frac{\partial^2 U(d_0(i))}{\partial d^2(i)} \times (\Delta d(i))^2$$

Where $\Delta d(i)$ is the customer demand change from $d_0(i)$ (the initial demand) to $d(i)$ (optimum point).

The customer benefit can take different units; however, in this paper, for the simplicity, it is assumed that this benefit is in terms of dollar. Moreover, according to the classic economics, it is assumed that every individual optimizes her benefit. To obtain the optimum point, the derivative of the benefit function with respect to the demand must be zero.

$$\frac{\partial B(d(i))}{\partial d(i)} = \frac{\partial U(d(i))}{\partial d(i)} - p(i) = 0 \quad (4)$$

Therefore,

$$\frac{\partial U(d(i))}{\partial d(i)} = p(i) \quad (5)$$

Hence, according to (5), in the optimum point, the marginal utility is equal to the price of the electricity.

Assuming that the initial demand before implementing the DR program is in optimum point, (6) and (7) should hold.

$$\frac{\partial B_0}{\partial d(i)} = \frac{\partial U(d_0(i))}{\partial d(i)} - p_0 = 0 \quad (6)$$

$$\frac{\partial U(d_0(i))}{\partial d(i)} = p_0 \quad (7)$$

By using (5) and the definition of the price elasticity of demand (1), (8) is obtained.

$$\frac{\partial^2 U(d(i))}{\partial d^2(i)} = \frac{\partial p}{\partial d} = \frac{1}{E} \times \frac{p_0}{d_0} \quad (8)$$

Plugging (7) and (8) into the Taylor series expansion of utility function (3) gives,

$$U(d(i)) = U(d_0(i)) + p_0.\Delta d(i) + \quad (9)$$

$$\frac{1}{2} \cdot \frac{1}{E(i)} \cdot \frac{p_0}{d_0} \cdot (\Delta d(i))^2$$

(9) could be rewritten and expanded as follows:

$$U(d(i)) = U(d_0(i)) + p_0.\Delta d(i)[1 + \frac{\Delta d(i)}{2 \times E(i) \times d_0}] \quad (10)$$

Expanding $\Delta d(i) = d(i) - d_0(i)$ and then plugging (10) into (5) yields (11) and (12),

$$p(i) = p_0(i) \times [1 + \frac{d(i) - d_0(i)}{E(i) \times d_0(i)}] \quad (11)$$

$$p(i) = p_0(i) + p_0(i) \times \frac{d(i) - d_0(i)}{E(i) \times d_0(i)} \quad (12)$$

Therefore, the customer's demand can be represented as follows:

$$d(i) = d_0(i) \times \left\{ 1 + \frac{E(i) \times (p(i) - p_0(i))}{p_0(i)} \right\} \quad (13)$$

*B. Demand-price elasticity model for 24 hours*

To provide a model for 24 hours, both self- and cross-elasticity have to be taken into the consideration. The cross-elasticity between hours *i* and *j* is defined as:

$$E(i,j) = \frac{P_0(j)}{d_0(i)} \times \frac{\partial d(i)}{\partial p(j)}, \quad i \neq j \quad (14)$$

The demand response model for 24 hours of a day could be obtained by combining self- and cross-elasticity of demand as follows:

$$d(i) = d_0(i) + E(i) \times \frac{d_0(i)}{p_0(i)} \times (p(i) - p_0(i)) + \quad (15)$$

$$\sum_{\substack{j=1 \\ j \neq i}}^{24} E(i,j) \times \frac{d_0(i)}{p_0(j)} \times (p(j) - p_0(j)), \quad i = 1, 2, ..., 24$$

The variation in the demand in (15) stems from two sources, one source is the self-elasticity which is reflected by the first and second terms and the other source is cross-elasticity which is reflected by the third term. The obtained relation in (15) is employed in this paper for demand response modeling part.

III. MULTIPLE LINEAR REGRESSION MODEL

As it was mentioned, to perform our proposed algorithm, the day-ahead price forecasting is necessary. An MLR model is used in this paper to carry out this task. This model is a linear model of demand, weather, time of day, week and season. In addition to the day-ahead price forecasting, this model could be utilized to update the spot market prices which is a part of our algorithm. Indeed, MLR function is able to model the dependent variable (price) as a linear function of the independent variables, independent dummy variables and interaction variables. The abundance of sample data can make the MLR model a very powerful tool.

Moreover, this model could be enhanced by adding different lag orders of the variables or different functional forms of the weather parameters; nevertheless, these extra variables are avoided in this paper for simplicity. Indeed, the accuracy of MLR model is enough for the purposes of this paper. The model used for the forecasting purpose is as follows:

$$P(i) = \alpha_0 + \sum_{k=1}^{24} \alpha_k . h_k(i) + \alpha_{25}.d(i) + \alpha_{26}.T(i) + \alpha_{27}.W(i) + \quad (16)$$

$$\alpha_{28}.M(i) + \alpha_{29}.D^H(i) + \alpha_{30}.D^{Sat}(i) + \alpha_{31}.D^{Sun}(i)$$

Where $\alpha_k$ is the coefficient of the independent variables, $h_k$ is hours of the day, $T(i)$ is the temperature at time $i$, $W(i)$ is the dew point at time $i$, $M(i)$ is the month at time $i$, $D^H(i)$ is a binary variable which indicates the holiday at time $i$, $D^{Sat}(i)$ and $D^{Sun}(i)$ are a binary variables which indicate the Saturdar and Sunday, respectively, at time $i$.

## IV. IMPLEMENTATION

Before explaining the implementation of the proposed algorithm, it is necessary to make several assumptions for this study. These assumptions are listed as follows.

- The utility is an independent non-profit agent that functions as an intermediary link between the customers and the wholesale market and purchases the electricity on behalf of the customers. However, in practice, the utility and the customers are two separate entities and have different profit functions.
- The spot market is the main market and the utilities purchase their whole demand in this market. However, in practice, the day-ahead market is the main market and the spot market is the real time market where the participants use it to meet their obligations in an emergency case.
- Real time pricing is applied to DR model. Indeed, to incent the customers to change their consumption pattern, real time pricing is the best possible choice.

To justify the aforementioned assumptions, several reasons could be provided. First, although considering the utility profit is more realistic, it makes the problem extremely complicated while does not provide any relevant outcome for this study. Second, the price volatility mainly exists in spot market where the limitation of supply leads to sharp price spikes which is the main focus of this study. However, in the day-ahead market, due to the more availability of supply, such sharp spikes are nonexistent. Indeed, by considering the spot market as the main market, it is expected to observe more pronounced results. Finally, the customers typically pay flat rate for electricity. However, in this paper, the real time pricing is selected to be applied to DR model as it would be the best incentive for the users to shift loads at different times.

By considering the aforementioned assumptions, the proposed algorithm for the evaluation of the impact of DR on spot market prices is illustrated in Fig. 2.

The procedure is as follows. First, the historical data including the information about price, demand, weather, time of day, week and season are loaded into the SAS software. Then the algorithm continues with developing a simple basic regression model. Since adding too many variables to the regression model may lead to the reduction of efficiency and accuracy, it is necessary to select the most efficient variables. Therefore, a simple basic model is used at the beginning and the others variables are added later on one at a time to evaluate whether or not it improves the out-of-sample forecasting error root mean square (FERMS).

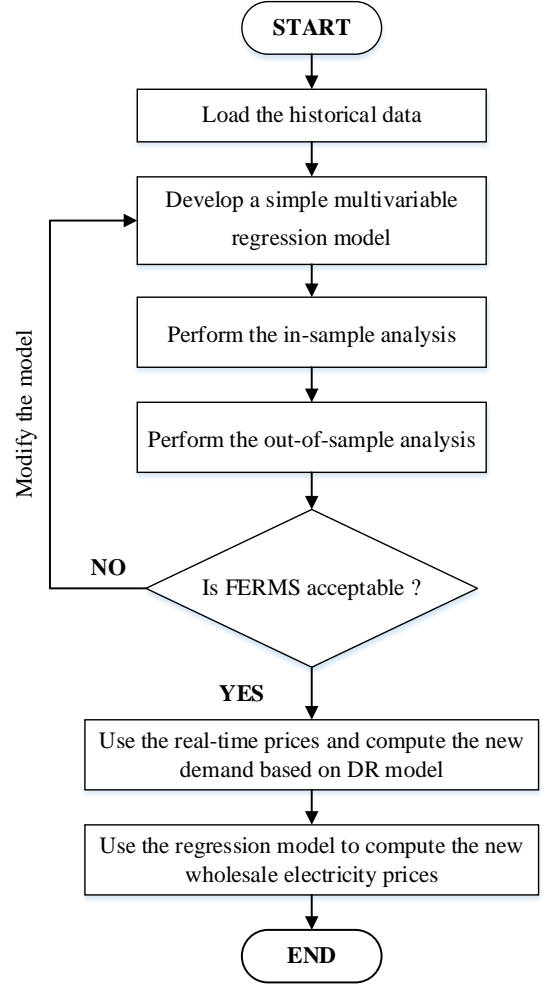

Fig. 2. flowchart illustrating the implementation process

TABLE I: SELF AND CROSS ELASTICITIES

|  | *Peak* | *Off-Peak* | *Low* |
|---|---|---|---|
| *Peak* | -0.10 | 0.016 | 0.012 |
| *Off-Peak* | 0.016 | -0.10 | 0.01 |
| *Low* | 0.012 | 0.01 | -0.10 |

If the overall FERMS is acceptable, then the day-ahead price forecasting could be performed. The resultant data could be fed into the selected DR model in a software. In this paper MATLAB is employed for this purpose. DR model which uses the day-ahead forecasted prices as a real time pricing produces a new demand. For implementing the DR model, the self- and cross- elasticity values of table I have been used [30-31]. Then by using the regression model and applying new demand, the updated prices could be achieved.

## V. CASE STUDY

TABLE II: MULTIVARIALBE REGRESSION MODEL
** 1% SIGNIFICANT LEVEL, * 5% SIGNIFICANT LEVEL, + 10% SIGNIFICANT LEVEL

| Variable | Parameter Estimate | t-value |
|---|---|---|
| Intercept | -182.3786 | -10.81** |
| Demand | 0.0483 | 20.66** |
| Temperature | 0.62756 | 2.63** |
| Humidity | -1.23893 | -7.22** |
| Month | 9.66399 | 8.41** |
| holiday | 7.23457 | 1.17 |
| Saturday | 9.30045 | 3.53** |
| Sunday | 11.82146 | 4.33** |
| hour1 | 15.86262 | 2.65** |
| hour2 | 24.07332 | 3.99** |
| hour3 | 27.62021 | 4.55** |
| hour4 | 28.27496 | 4.64** |
| hour5 | 28.42 | 4.67** |
| hour6 | 24.2956 | 4.02** |
| hour7 | 11.63418 | 1.93+ |
| hour8 | 0.09629 | 0.02 |
| hour9 | -12.40514 | -2.05* |
| hour10 | -23.12766 | -3.79** |
| hour11 | -29.87756 | -4.82** |
| hour12 | -32.4509 | -5.14** |
| hour13 | -34.80398 | -5.43** |
| hour14 | -29.86742 | -4.57** |
| hour15 | -35.05025 | -5.32** |
| hour16 | -28.77225 | -4.34** |
| hour17 | -23.27377 | -3.5** |
| hour18 | -35.94987 | -5.45** |
| hour19 | -45.16361 | -6.97** |
| hour20 | -39.38583 | -6.21** |
| hour21 | -35.50788 | -5.67** |
| hour22 | -29.14752 | -4.73** |
| hour23 | -15.38141 | -2.56* |

### A. Description

The proposed approach is examined on the reported hourly data of New England ISO [32]. This ISO provides the zonal information for all its serving areas including the day-ahead price, load forecasting, real locational marginal price (LMP), load, Dry Bulb temperature and Dew Point. For this study, the data of summer 2014 in Connecticut is used. The study has been done for one week (August 18th to 24th).

Moreover, in this case study, it is assumed that the customers had paid a flat rate of 30$/MWh for this week before DR program. After DR program, the utility charges the wholesale market LMP for each hour.

### B. Results and discussion

After applying the proposed approach to the selected data and several iterations, variables in table II are selected as the most efficient variables for the forecasting purposes. Table II lists the variables, their parameter estimates (coefficients) and their t-value.

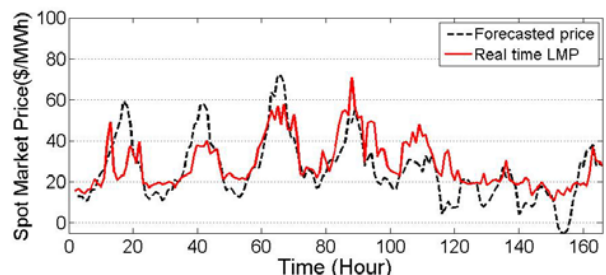

Fig. 3. forecasted price vs. real price

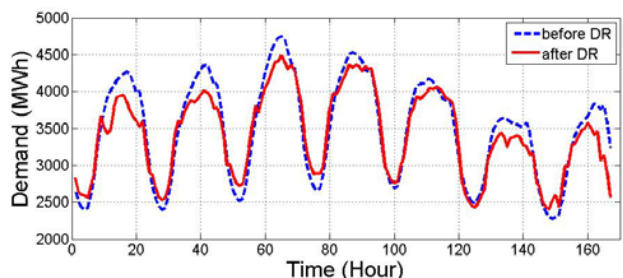

Fig. 4. Demand before and after DR program

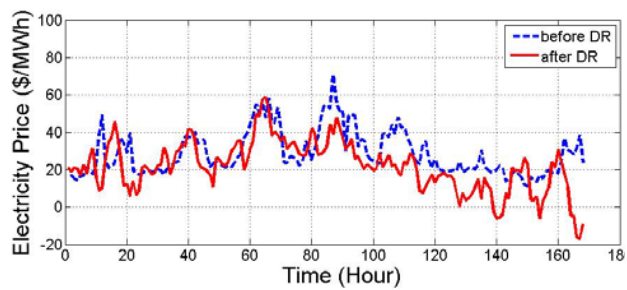

Fig. 5. spot market price before and after DR program

TABLE III: SUMMARY OF THE EFFECT OF DR ON THE SPOT MARKET PRICE

| Change in Energy | -13058 MWh | -2.23 % |
|---|---|---|
| Change in Cost | -4,674,396 $ | -26.23 % |

The parameter estimate could be interpreted as the change in the predicted value of the dependent variable (price) for one unit increase in the independent variable. Also, t-value is defined as a ratio of the departure of an estimated parameter from its notional value and its standard error. In table II, with 31 degrees of freedom (number of variables and intercept), t-values between 1.3 and 1.69 are significant with 10% error. t-values between 1.69 and 2.45 are significant with 5% error and t-values over 2.45 are significant with 1% error. Therefore, from table II, it is understood that almost all of the variables are significant with 1% error. However, the selection of some variables with lower level of significance is necessary in terms of improving the out-of-sample analysis and FERMS.

By using the coefficients of table II and comparing the results with the real time LMP, the forecasting error root mean square (FERMS) of electricity price for this case study is computed as 11% which is an acceptable error for volatile variable like electricity price. Fig. 3 illustrates the forecasting prices and real time LMP.

The achieved forecasted prices are fed into the selected DR model to produce the new demand. Fig. 4 shows the demand before and after applying the demand response model. Finally, the new demand is fed into the regression model to produce new prices.

Fig. 5 shows the wholesale electricity price before and after applying the DR model. As it is shown, the price spikes are declined considerably.

Based on the demand and price before and after applying DR model, the total pertinent cost of electricity could be computed. According to the calculation, it is observed that the total demand is reduced by 2.23% and the total cost of electricity for this week is reduced by 26.23%. Table III summarizes the change in the wholesale electricity cost.

26.23% change in the total cost stems from two major sources. First, 2.23% change in the total demand; second, a considerable shift of demand from expensive peak time to less-expensive off-peak period of time due to customer's exposure to the real time pricing as discussed earlier.

## VI. CONCLUSION

This paper studied the effect of demand response programs on the electricity spot market price. Demand response model was used to account for the customer reaction function facing the real time pricing. MLR model was employed to perform the price forecasting in day-ahead market. Then, an algorithm comprised of MLR and DR model was introduced to feed the outcome of DR model into the spot market and explore the impact of demand response program. Finally, the proposed approach was tested on a case study by using the real data of New England ISO.

The results showed that the demand-side participation through 2.23% reduction of demand as well as shifting load from expensive peak time to less-expensive off-peak period of time could reduce the total cost of electricity by approximately 26 percent.

Therefore, according to the results, it could be said that keeping the demand side isolated from the market deprives the market of an effective tool for smoothing out the price volatility. However, with adopting the proper DR program, both the supply- and demand-side of the market can benefit. The customers can benefit from the reduced bill and the supply-side can compensate shortage of its sources during the peak time.